\documentclass[aps,prl,twocolumn,showpacs,superscriptaddress,10pt]{revtex4-1}

\usepackage{graphicx}
\usepackage{amsmath,amssymb,amstext,amsthm}
\usepackage{bbold}
\usepackage{cancel}
\usepackage{color}
\usepackage{verbatim}
\usepackage{booktabs}
\usepackage[caption=false]{subfig}
\usepackage[english]{babel}

\newcommand{\Z}{\mathbb{Z}}

\usepackage[utf8x]{inputenc}
\begin{document}

\title{Synthesizing Weyl semimetals in weak topological insulator and \\ topological crystalline insulator multilayers}

\author{Alexander Lau}
\affiliation{Institute for Theoretical Solid State Physics, IFW Dresden, 
01171 Dresden, Germany}

\author{Carmine Ortix}
\affiliation{Institute for Theoretical Solid State Physics, IFW Dresden, 
01171 Dresden, Germany}
\affiliation{Institute for Theoretical Physics, Center for Extreme Matter and
Emergent Phenomena, Utrecht University, Princetonplein 5, 3584 CC Utrecht,
Netherlands}

\date{\today}

\begin{abstract}
We propose a different route to time-reversal invariant Weyl semimetals
employing multilayer heterostructures comprising ordinary ``trivial" insulators and nontrivial insulators with \textit{pairs} of protected Dirac cones on the surface.
We consider both the case of weak topological insualtors, where surface Dirac cones are pinned to time-reversal invariant momenta, and of topological crystalline insulators with unpinned surface Dirac cones.
For both realizations we explain phenomenologically how the proposed construction leads to the emergence of a Weyl semimetal phase. We further formulate effective low-energy models for which we prove the existence of semimetallic phases with four isolated Weyl points. Finally, we discuss how the proposed design can be realized experimentally with state-of-the-art technologies.
\end{abstract}

\maketitle

\paragraph{Introduction -- }

Topological phases are novel states of matter whose study has led to a plethora of fascinating discoveries and developments in modern condensed-matter physics~\cite{QiZ11,HaK10,KDP80,TKN82,SaA16,MZF12,Bur16,Zha92,Sen15,DXG16,LOB15_2}. 
In gapped systems, the quantized invariant of a topological quantum state of matter is directly 
related to the presence of protected edge or surface states by the so-called bulk-boundary correspondence~\cite{QiZ11,HaK10}. The most famous examples of topological materials are two-dimensional (2D) and three-dimensional (3D) time-reversal invariant (TRI) topological insulators (TIs)~\cite{KaM05,FKM07,BHZ06,KWB07,HXW09,XQH09,RIR13,PRK15,LFY11}, as well as topological crystalline insulators (TCIs)~\cite{AnF15,Fu11,HLL12,TRS12,AFG14,LBO16}.

Weyl semimetals (WSMs), instead, are members of the family of \emph{gapless} topological phases~\cite{Bur16,YZT12,YoK15,KKE16}, and have recently been discovered experimentally~\cite{HXB15,LWF15,LXW15,XBA15,XAB15,WFF15,HKE17}. WSMs are 3D materials whose bulk energy bands cross linearly at isolated points in the Brillouin zone (BZ),  the so-called Weyl nodes~\cite{WTV11,BuB11,ZWB12,Oja13}. Around these points the system can effectively be described by a Weyl Hamiltonian of the general form $H(\mathbf{k})=\sum_{ij} k_iA_{ij}\sigma_j$,
where $i=x,y,z$, $j=0,x,y,z$ and $\sigma_j$ are Pauli matrices~~\cite{SGW15}. A necessary condition for a WSM is the absence of either time-reversal or inversion symmetry since the simultaneous presence of both requires any band crossing to be at least a four-fold degenerate Dirac point.

Weyl nodes represent monopoles of the Berry flux in momentum space, and can be
assigned a well-defined chirality or topological charge. They always come in pairs of opposite chirality due to the charge neutrality of the BZ. 
Moreover, in the case of TRI Weyl semimetals,
the minimal number of Weyl points is four since time reversal always connects Weyl points with same chirality. Perturbations can merely shift the nodes in energy or momentum. Therefore, Weyl points are stable bulk features~\cite{BuB11,OkM14}. Furthermore, WSMs host robust surface states, commonly referred to as Fermi arcs, connecting Weyl points with opposite topological charge~\cite{WTV11,LKB17}. In addition, it has been shown recently that these characteristic arc features can coexist with topological Dirac cones on the surface of a Weyl semimetal~\cite{LKB17,JuT17,TSG17}.

In this Rapid Communication, we present a multilayer design for a TRI Weyl semimetal. Multilayer heterostructures have been proposed for inversion-symmetric~\cite{BuB11} and also for time-reversal symmetric WSMs~\cite{HaB12}, but only considering strong TIs with a single Dirac cone per surface
as the active layer. We extend this principle to topological materials with an even number of surface Dirac cones. In particular, we consider two distinct cases: multilayers based on weak TIs with two Dirac cones pinned to TRI momenta, and multilayers based on TCIs with two unpinned Dirac cones. We show that both systems give rise to TRI Weyl-semimetal phases with four isolated Weyl nodes. We also find strong TI and weak TI phases in the multilayer phase diagrams, which renders the proposed designs a suitable platform for 
the artificial synthesis of 3D TIs.
Moreover, we discuss 
the experimental feasibility of our theoretical proposal.

\paragraph{Weak TI multilayer -- }

TRI topological insulators are realizations of nontrivial
topological phases in the Altland Zirnbauer class AII~\cite{Zir96,AlZ97,RSF10}. In three dimensions,
gapped systems belonging to this class
can be characterized by the four
$\Z_2$ topological invariants $\nu_0;(\nu_1\nu_2\nu_3)$~\cite{FKM07,FuK06}.
Insulators with nonzero $\nu_0$ are called strong TIs.
Their hallmark is the existence of an odd number of protected surface Dirac cones \emph{pinned} to TRI momenta~\cite{FKM07}.

If the strong index $\nu_0$ is zero but at least one of the weak indices $\nu_1$,
$\nu_2$, $\nu_3$ is 
non-zero,
the system is dubbed a weak TI. In contrast to their strong relatives, weak TIs feature an even
number of topologically protected, \emph{pinned} Dirac cones only at certain surfaces~\cite{FKM07,LOB15}. More
specifically, there exist so-called ``dark surfaces'' 
where surface Dirac cones are absent.
These are the
surfaces whose Miller indices (modulo 2) are identical to the weak indices
$(\nu_1\nu_2\nu_3)$. This property is related to the fact that a weak TI is
topologically equivalent to a stack of 2D TIs with stacking
direction $[\nu_1\nu_2\nu_3]$. 

\begin{figure}[t]\centering
\includegraphics[width=1.0\columnwidth]
{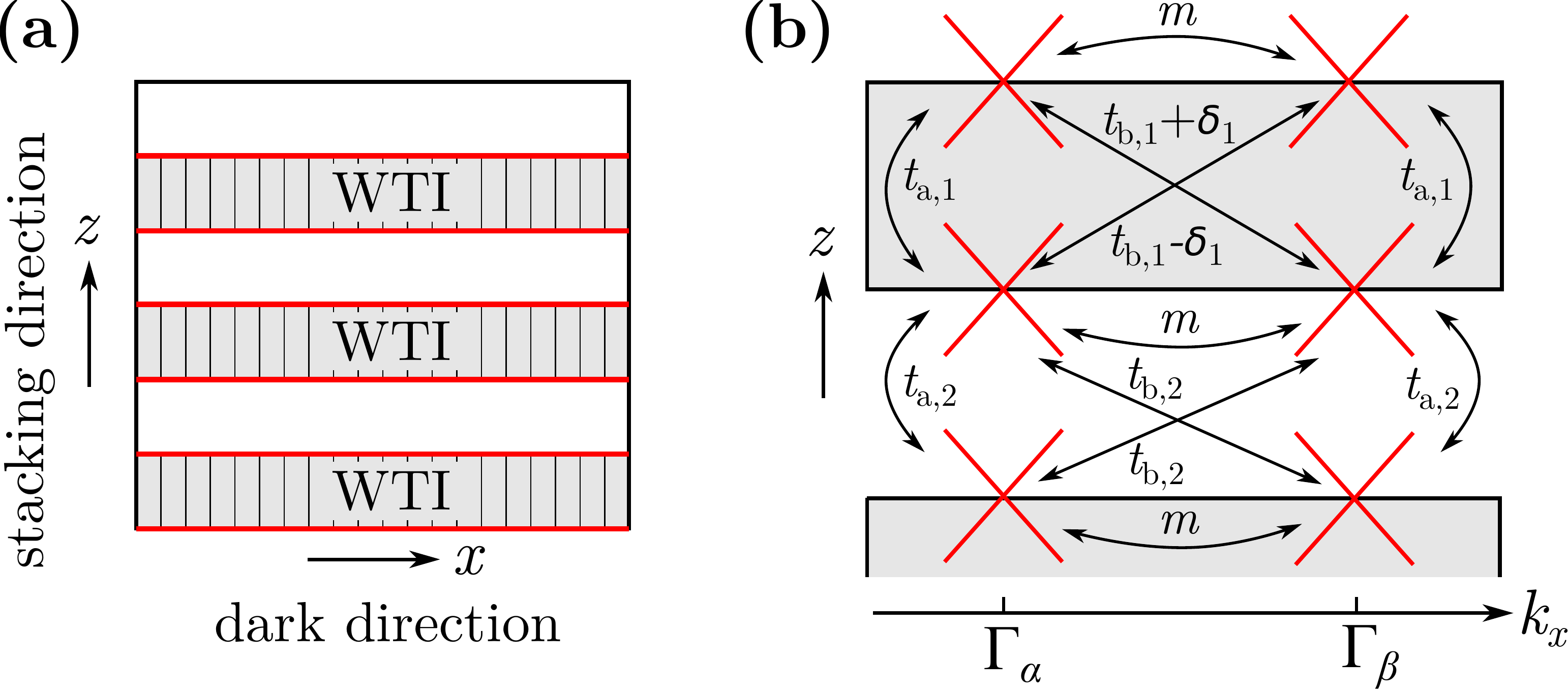}
\caption{(color online) Weak TI multilayer: (a) cartoon
of the multilayer design. The surfaces of the weak TI layers, which contribute two pinned Dirac cones each, are highlighted in red. The layered
structure of the weak TIs is indicated. (b) Schematic of the coupling
terms between the surface Dirac cones (red). The Dirac cones are pinned to different TRI momenta
$\Gamma_\alpha$ and $\Gamma_\beta$ which are mapped onto each other by translational-symmetry breaking induced by the dimerization mass $m$.}
\label{fig:WTI_heterostructure}
\end{figure} 

Let us now consider a heterostructure consisting of layers of
weak TIs as illustrated in Fig.~\ref{fig:WTI_heterostructure}(a). Without loss of generality, we consider the invariants of the weak TIs to be $0;(100)$, i.e., the weak TIs are equivalent to 2D TIs stacked in the $x$ direction,
which thus corresponds to
the ``dark direction''. Next, 
we create a one-dimensional superlattice in the
$z$ direction, which is perpendicular to the dark direction, by
inserting
spacers of ordinary insulators (OIs) between the weak TI layers.
Due to the bulk-boundary correspondence, there will be an even number of Dirac cones at
each interface between the OI and the weak TI.
For simplicity, we here assume each interface to have the minimal number of two Dirac cones. Initially, the Dirac cones are pinned to \emph{different} TRI momenta $\Gamma_\alpha$ and $\Gamma_\beta$ along the dark direction. 
However, a dimerization in the weak TI crystal~\cite{MBM12} breaks the translational symmetry in the dark direction $x$ and allows the two Dirac cones to couple.
Furthermore, if weak TI layers and spacer layers are sufficiently thin, also Dirac states from adjacent surfaces can couple through hybridization.

The low-energy theory of the multilayer heterostructure is then effectively described by the following Hamiltonian,
\begin{eqnarray}
\mathcal{H} &=& \sum_{\mathbf{k}_\perp,ij}\Big[v_D
\sigma^3\tau^0(\hat{z}\times\mathbf{s})\cdot\mathbf{k}_\perp\,\delta_{i,j}
+ m\sigma^0\tau^2 s^3\, \delta_{i,j}\nonumber\\
&&{}+ t_\textrm{a,1} \sigma^1\tau^0 s^0\, \delta_{i,j}
+ \frac{t_\textrm{a,2}}{2} (\sigma^{-}\,\delta_{i,j+1} +
\sigma^{+}\,\delta_{i,j-1})\tau^0 s^0 \nonumber\\
&&{}+ t_\textrm{b,1} \sigma^1\tau^1 s^0\,\delta_{i,j} + 
\delta_1 \sigma^2\tau^2 s^0\,\delta_{i,j}\nonumber\\
&&{}+ \frac{t_\textrm{b,2}}{2}(\sigma^{-}\,\delta_{i,j+1} +
\sigma^{+}\,\delta_{i,j-1}) \tau^1 s^0\,\Big]\, c_{\mathbf{k}_\perp,j}^\dagger
c_{\mathbf{k}_\perp,i},
\label{eq:WTI_multilayer_Hamiltonian}
\end{eqnarray}
where $\sigma^\nu$, $\tau^\nu$, and $s^\nu$ are Pauli
matrices associated with the top and bottom surfaces of the weak TIs, the
two Dirac cones per surface, and the spin degree of freedom, respectively.
In addition,
$v_D$ is the Fermi velocity of the Dirac fermions, $\mathbf{k}_\perp=(k_x,k_y)$
is their momentum in the 2D surface BZ of the weak TIs, $m$ is the ``dimerization mass'', 
and the indices $i$, $j$ label the weak TI layers. The parameters of our model are illustrated in Fig.~\ref{fig:WTI_heterostructure}(b). Note that we also allow for an inter-valley coupling imbalance $\delta_1$. Such an imbalance is expected to arise naturally when top and bottom surfaces of the weak TI layers are nonidentical, 
{\it e.g.} when one surface is canted relative to the other.
For simplicity, other imbalances have been omitted since they do not change our results qualitatively. 

The model of Eq.~\eqref{eq:WTI_multilayer_Hamiltonian} preserves time-reversal symmetry with the operator $\Theta = i\sigma^0\tau^0
s^2\,K$, $\mathbf{k}\rightarrow -\mathbf{k}$, where $K$ is complex
conjugation. The inter-valley coupling imbalance serves as an inversion-symmetry breaking term. The corresponding inversion operator is $P=\sigma^1\tau^0 s^0$, with $\mathbf{k}\rightarrow -\mathbf{k}$.
In order to pin down the existence of a WSM phase in our model, we compute the energies and monitor the half-filling gap 
of the system in its parameter space. This allows us to identify
gap closing points and their degeneracies. 
We can anticipate the existence of a WSM phase qualitatively using the following arguments:
the hybridization of the surface Dirac cones $\propto t_{a, 1}, t_{a, 2}$ leads to two 3D Dirac points in the 3D BZ of the model. The dimerization $\propto m$ can either gap out these degeneracy points,
leading to fully gapped phases, or shift the two 3D Dirac points in momentum space. Such an unstable Dirac semimetal phase can in principle be transformed into a stable WSM phase by
breaking inversion symmetry~\cite{BuB11,ZWB12,Oja13}.
This is accomplished by the inter-valley coupling imbalance: each 3D Dirac point is split into two separate Weyl points.

\begin{figure}[t]\centering
\includegraphics[width=1.0\columnwidth]
{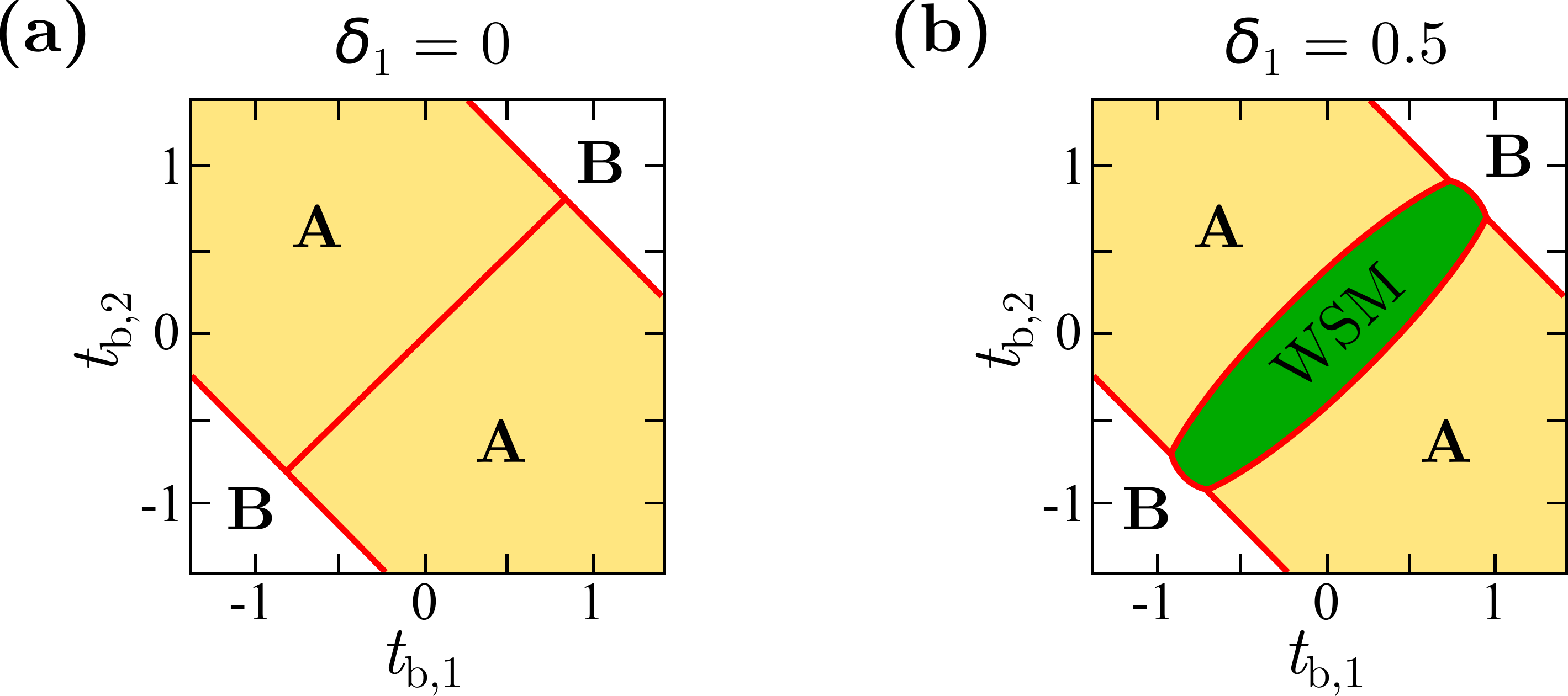}
\caption{(color online) Phase diagrams of the weak TI multilayer with
$v_D=m=t_\mathrm{a,1}=t_\mathrm{a,2}=1$. (a) With inversion symmetry ($\delta_1=0$): there are two different gapped phases, phase A (strong TI) and phase B (OI), and an unstable Dirac semimetal phase (red lines). (b) Broken inversion symmetry ($\delta_1=0.5$): between the strong TI phases, a stable WSM phase (green) with four separate Weyl points emerges.}
\label{fig:phase_diagrams}
\end{figure} 

Let us first explore the half-filling phase diagram for the inversion-symmetric model with $\delta_1=0$ [see
Fig.~\ref{fig:phase_diagrams}(a)]. We find several gapped phases, which we dub A and B, separated by
phase boundaries along which the system is semimetallic. More specifically,
in the semimetallic phase the system exhibits two unpinned 3D Dirac points on the $k_z$
axis related by time-reversal symmetry.

By exploring the parameter space of our model, we find that the B phases are adiabatically connected to a multilayer of fully-decoupled dimerized weak TIs. Since a dimerized weak TI is topologically trivial, these phases correspond to an ordinary insulator with $\Z_2$ invariants $0;(000)$. In order to determine the nature of the other gapped phases, we calculate the corresponding parity eigenvalues of all occupied states at the TRI momenta $\Gamma=(0,0,0)$ and $Z=(0,0,\pi)$~\cite{FuK07}, assuming that the topologically active band inversions occur only at these momenta. 
We find that a band inversion at the $\Gamma$ point 
occurs by moving from a
B phase to an A phase. On the contrary, no band inversion occurs between the two A phases. Hence, under our assumption, we deduce that the A phases correspond to strong TIs with $\Z_2$ invariants $1;(000)$. We have confirmed these findings by analyzing a lattice regularization of our model (see Supplemental Material). 

Let us now turn to the inversion-broken case ($\delta_1\neq 0$). With inversion symmetry, the two strong TI phases were separated by a Dirac-semimetal line. By breaking inversion symmetry, the Dirac points
are split into four separate Weyl points along the $k_z$ axis. In this way, a WSM  
stability region
emerges in
the phase diagram [Fig.~\ref{fig:phase_diagrams}(b)]. By integrating the Berry
curvature over a closed momentum-space surface around each of the Weyl nodes, we calculate their
topological charges to be $\pm 1$.

In order to synthesize our proposed weak TI multilayer, one could start out with a dimerized weak TI material, namely Bi$_{13}$Pt$_3$I$_7$~\cite{PRK15}. 
Into the dark surface of the material one could then 
carve an array of sufficiently thin channels, where the channels are alternatingly tilted against each other. In this way, opposite weak TI surfaces are nonidentical thereby providing the required interlayer coupling imbalance. This setup is extremely challenging but can be accomplished 
using terraces or creating trenches by, e.g., focused ion beams.
The coupling between the layers can be fine-tuned by varying the channel spacing, the channel width, and their relative angle. Finally, the channels must be filled with an insulating spacer material. The resulting sample can
be viewed as a dimerized weak TI with a WSM layer on top of it. 
The characteristic features of the WSM, such as surface Fermi arcs, could be then detected performing angle-resolved photoemission experiments. 

\paragraph{TCI multilayer -- }

The essential ingredient 
used in the setup above is the presence of two Dirac cones per surface
which are coupled to each other. 
The question that arises is whether this idea is also applicable to systems in which surface Dirac cones are not pinned to TRI momenta. This occurs, for instance, in
the recently discovered TCIs in the SnTe material class, which allow for an even number of \emph{unpinned} surface Dirac cones protected by mirror symmetry~\cite{HLL12}.

TCIs are similar to ``conventional" TIs except that topological states are protected by additional discrete symmetries, such as mirror, inversion or space group symmetries~\cite{Fu11,CYR13,ShS14,HPB11,LBO16,SMJ12}. For systems in the SnTe material class, 
this can be understood as follows:
in the BZ of a mirror-symmetric material, there are planes that are invariant under the mirror operation $M$.
Since the Bloch Hamiltonian $H(\mathbf{k})$ commutes with $M$ in these planes, $H(\mathbf{k})$ can be brought in block form with respect to the mirror eigenvalues $\pm i$ of its eigenstates. For a mirror-invariant plane, we can assign a Chern number $n_{\pm i}$ to each of the two blocks. Furthermore, one can show that $n_{+i} = -n_{-i}$. Hence, $n_M = (n_{+i}-n_{-i})/2$ defines a $\Z$ topological invariant, the so-called mirror Chern number. In this sense, a TCI is a material with a nonzero mirror Chern number. By bulk-boundary correspondence, a nonzero mirror Chern number implies the presence of $|n_M|$ Dirac cones on surfaces which preserve the protecting mirror symmetry $M$~\cite{HLL12}.

\begin{figure}[t]\centering
\includegraphics[width=1.0\columnwidth]
{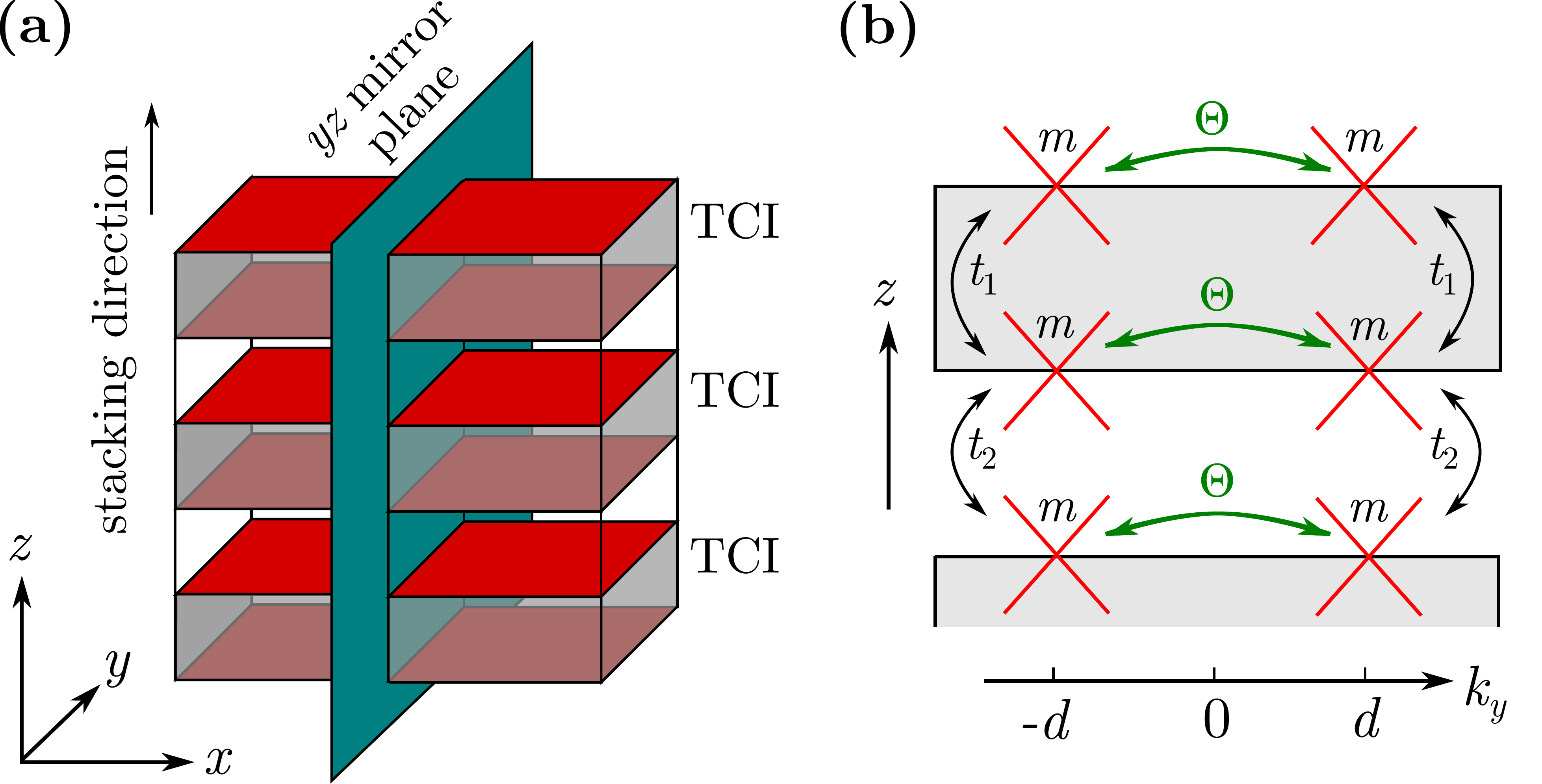}
\caption{(color online) TCI multilayer: (a) cartoon of the multilayer design. The surfaces of the TCI layers (red) contribute two unpinned surface Dirac cones each which are protected by a $yz$ mirror plane. (b) Schematic of the coupling terms between the surface Dirac cones (red). The Dirac cones are connected by time reversal $\Theta$ and can be gapped by a mirror-symmetry breaking mass $m$.}
\label{fig:TCI_multilayer_schematics}
\end{figure} 
In the remainder of this paper,
we are going to show that also TCI multilayers give rise to WSM phases.
In analogy with weak TI heterostructures, let us consider a multilayer consisting of alternating layers of TCIs and OIs [see Fig.~\ref{fig:TCI_multilayer_schematics}(a)]. 
Moreover, we will
use a minimal TRI TCI with mirror Chern number $n_M=2$ protected by a $yz$ mirror plane. Thus, on surfaces parallel to the $xy$ plane there will be two \emph{unpinned} Dirac cones 
at momenta related by time-reversal symmetry.
Without loss of generality, let them be at $\mathbf{k}_\perp = \pm\mathbf{d}=(0,\pm d)$. As opposed to 
a dimerized weak TI multilayer, the Dirac cones can be gapped even in the absence of intravalley scattering by
breaking mirror-symmetry with respect to the $yz$ mirror plane. This can be accomplished, for instance, by a ferrorelectric distortion~\cite{HLL12,VeF16}. Moreover, let the stacking direction of the TCI layers coincide with the $z$ axis. 

Taking into account a ferroelectric Dirac mass parameterized by $m$, which breaks the mirror symmetry of the system, the low-energy theory of the TCI multilayer can be written down in analogy with the weak TI heterostructure. The model parameters are illustrated in Fig.~\ref{fig:TCI_multilayer_schematics}(b).   
The corresponding Hamiltonian is
\begin{eqnarray}
\mathcal{H} &=& \sum_{\mathbf{k}_\perp,ij}\Big[\frac{v_D}{2}
\sigma^3(\tau^0 + \tau^3)(\hat{z}\times\mathbf{s})\cdot(\mathbf{k}_\perp+\mathbf{d})\,\delta_{i,j}\nonumber\\
&&{}+ \frac{v_D}{2}
\sigma^3(\tau^0 - \tau^3)(\hat{z}\times\mathbf{s})\cdot(\mathbf{k}_\perp-\mathbf{d})\,\delta_{i,j}\nonumber\\
&&{}+ t_1 \sigma^1\tau^0 s^0\, \delta_{i,j}
+ \frac{t_2}{2} (\sigma^{-}\,\delta_{i,j+1} +
\sigma^{+}\,\delta_{i,j-1})\tau^0 s^0 \nonumber\\
&&{}+ m\,\sigma^0\tau^3 s^3\, \delta_{i,j}\,\Big]\, c_{\mathbf{k}_\perp,j}^\dagger
c_{\mathbf{k}_\perp,i}. 
\label{eq:TCI_multilayer_Hamiltonian}
\end{eqnarray}
The reflection operator is $M_x=i\sigma^0\tau^0 s^1$ with $k_x \rightarrow -k_x$. Time-reversal symmetry is preserved for all parameters with 
$\Theta = i\sigma^0\tau^1 s^2\,K$, and $\mathbf{k}\rightarrow -\mathbf{k}$. The operator of spatial inversion is represented by $P=\sigma^1\tau^1 s^0$ (with $\mathbf{k}\rightarrow -\mathbf{k}$), and does not commute with the ferroelectric Dirac mass. Hence, the ferroelectric distortion also breaks inversion symmetry which will enable us to create a stable WSM phase similar to the weak TI multilayer.

\begin{figure}[t]\centering
\includegraphics[width=1.0\columnwidth]
{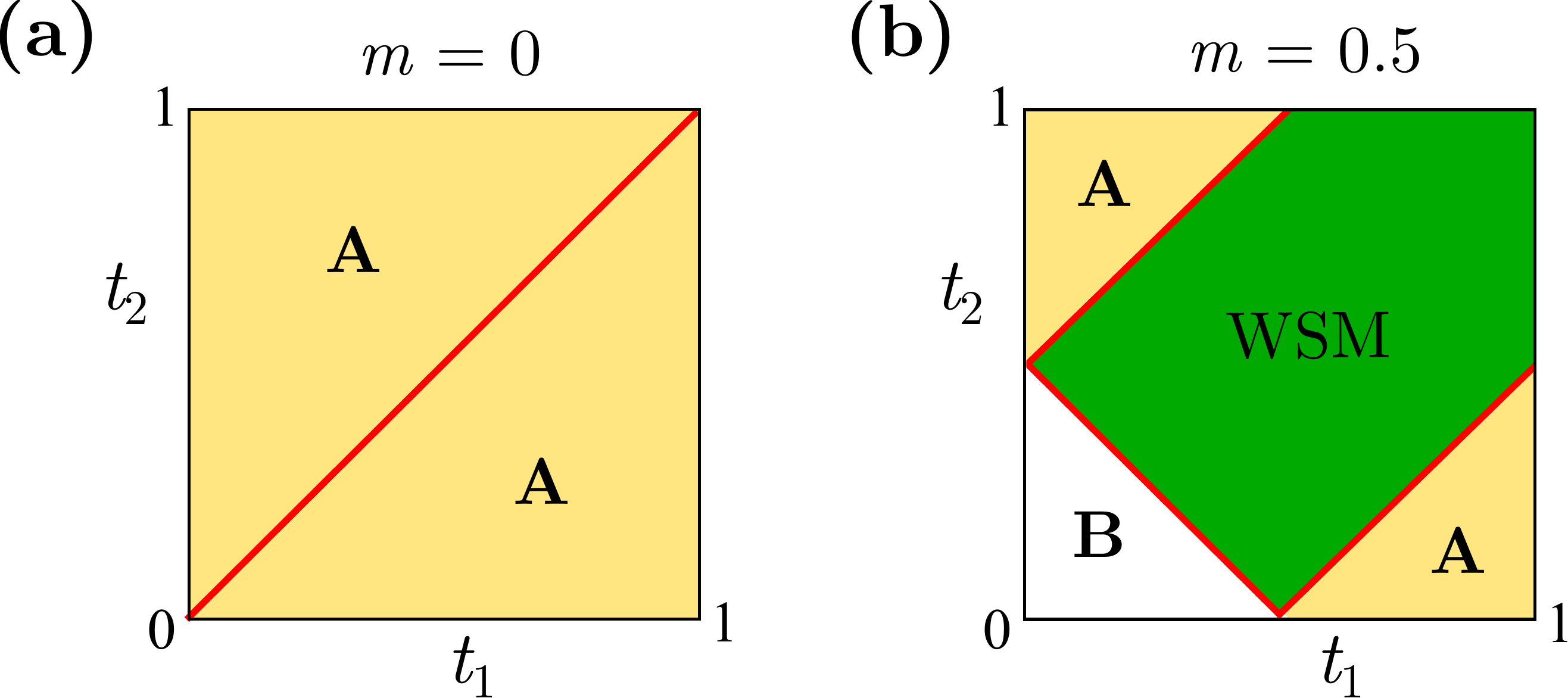}
\caption{(color online) Phase diagrams of the TCI multilayer with
$v_D=1$ and $\mathbf{d}=(0,1)$. (a) With inversion symmetry ($m=0$): there are two gapped phases A (weak TI) and an unstable Dirac semimetal phase (red line). (b) Broken inversion symmetry ($m=0.5$): a new gapped phase B (OI) emerges. Between the gapped phases there is a stable WSM phase (green) with four isolated Weyl points.}
\label{fig:TCI_multilayer_phase_diagrams}
\end{figure} 
In Figs.~\ref{fig:TCI_multilayer_phase_diagrams}(a) and~(b), we show the $t_1$-$t_2$ phase diagrams of the heterostructure. For $m=0$, the system preserves inversion symmetry. In this case, we find gapped phases for $t_1 \lessgtr t_2$, as well as a gapless phase along the line $t_1=t_2$ as is shown in Fig.~\ref{fig:TCI_multilayer_phase_diagrams}(a). In the latter, the system has two isolated bulk Dirac points at $\mathbf{k}=(0,\pm d,\pi)$. 

For nonzero $m$, a WSM phase with four isolated Weyl points emerges in the phase diagram [see Fig.~\ref{fig:TCI_multilayer_phase_diagrams}(b)]. The Weyl nodes have topological charges of $\pm 1$. Furthermore, a new gapped phase appears around $(t_1,t_2)=(0,0)$. 

Let us briefly comment on the gapped phases in Figs.~\ref{fig:TCI_multilayer_phase_diagrams}(a) and~(b). For $m\neq 0$, the point $(0,0)$ in the phase diagram represents a multilayer of decoupled, mirror-symmetry broken TCIs with gapped Dirac cones. This is a topologically trivial system. Hence, the B phase corresponds to a phase of OIs with $\Z_2$ invariants $0;(000)$. To determine the nature of the A phases, let us assume our model describes the low-energy theory of a lattice model. By analyzing how the parities change in the inversion-symmetric case ($m=0$) by going from one A phase to the other, we find that both A phases are identical. For the inversion-symmetry broken case, let us consider the transition from A phase to B phase along the $t_1$ or along the $t_2$ axis. We observe that the bulk energy gap of the system closes along lines in momentum space at $(k_x,k_y)=(0,\pm d)$. Such a gap closing transition can only change the weak $\Z_2$ invariant $\nu_3$ relative to the B phase while the others remain unchanged. Hence, the A phases correspond to weak TIs with $\Z_2$ invariants $0;(001)$, i.e., their dark direction coincides with the stacking direction of the multilayer. We have verified this conclusion
by comparing our results to a lattice regularization of our model (see Supplemental Material).

As opposed to the weak TI multilayer, the heterostructure based on TCI layers could be prepared 
in the form of a superlattice.
The coupling between the surface Dirac cones can be adjusted by choosing a particular thickness for the layers of TCIs and OIs. To break mirror symmetry on the surfaces of the TCI layers, one could use an insulating, ferro-electric material for the layers in between the TCIs. 

A concrete material candidate is a heterostructure of alternating layers of PbTe and SnTe stacked in the [110] direction~\cite{HLL12}. SnTe is a TCI and hosts two Dirac cones on (110) surfaces protected by a $(1\bar{1}0)$ mirror plane, whereas PbTe is an OI. Moreover, SnTe undergoes a ferroelectric distortion at low temperatures~\cite{HLL12}. The distortion is along the [111] direction. This breaks the $(1\bar{1}0)$ mirror plane thereby  providing an intrinsic mechanism to gap out the surface Dirac cones. Hence, this superlattice is expected to realize a WSM phase. 

\paragraph{Conclusions -- }

We have demonstrated that multilayer heterostructures based either on weak topological insulators or on topological crystalline insulators represent a novel platform for the study of time-reversal invariant Weyl semimetals. In the proposed designs, thin layers of the materials are stacked on top of each other while inserting spacer layers of ordinary insulators in between. At the interfaces, pairs of pinned or unpinned Dirac cones, which are coupled to each other, provide the main ingredient of the multilayer designs. Weyl phases are stabilized by breaking inversion symmetry either by canting the interfaces or by a ferroelectric distortion.

We have shown that both design principles give rise to stable Weyl semimetal phases with four isolated Weyl points in the Brillouin zone. Moreover, the phase diagrams also indicate the possibility of strong topological insulator phases in the weak topological insulator multilayer, and of weak topological insulator phases in the topological crystalline insulator heterostructure. As a result, the multilayers may also provide a different way of designing artificial 3D topological insulators. Finally, we have given realistic pathways on how to prepare the proposed heterostructures by using available materials, like Bi$_{13}$Pt$_3$I$_7$ or SnTe, and state-of-the-art technologies.

We acknowledge the financial support of the Future and Emerging Technologies (FET) programme within
the Seventh Framework Programme for Research of the European Commission 
under FET-Open grant number: 618083 (CNTQC).  
C.O. acknowledges support from the Deutsche Forschungsgemeinschaft (Grant No. OR 404/1-1), and from a VIDI grant (Project 680-47-543) financed by the Netherlands Organization for Scientific Research (NWO).

%\clearpage

%\newpage

%%%%%%%%%%%%%%%%%%%%%%%%%%%%%%%%%%%%%%%%%%%%%%%%%%%%%%%%%%%%%%%%%%%%%%%%%%%%%%%%%%%%%%%%%%%%%%%
%%%%%%%%%%%%%%%%%%%%%%%%%%%%%%%%% SUPPLEMENTAL MATERIAL %%%%%%%%%%%%%%%%%%%%%%%%%%%%%%%%%%%%%%%
%%%%%%%%%%%%%%%%%%%%%%%%%%%%%%%%%%%%%%%%%%%%%%%%%%%%%%%%%%%%%%%%%%%%%%%%%%%%%%%%%%%%%%%%%%%%%%%

\section*{SUPPLEMENTAL MATERIAL}

\section*{A: Tight-binding model for weak topological insulator multilayers}

In this section, we 
construct a tight-binding model whose low-energy theory is represented
by the Hamiltonian introduced in Eq.~(1) of the main part of the paper, and, for simplicity, we regularize the continuum model on a cubic lattice. 

We start by taking the momentum-space
Hamiltonian corresponding to Eq.~(1) of the main part of the paper, and perform
the replacements $k_{x,y}\rightarrow \sin k_{x,y}$. This 
yields a Weyl
semimetal with 16 Weyl points. In order to get a model with the minimal number
of Weyl points allowed by symmetry, which 
corresponds to  
four for a time-reversal invariant system, we can
annihilate all Weyl points away from the $k_z$ axis. This is accomplished by
replacing $m$ with $m+b(2-\cos k_x - \cos k_y)$, where we have introduced 
the additional tight-binding parameter $b$. The Bloch Hamiltonian of the ensuing model reads
\begin{eqnarray}
H(\mathbf{k}) &=& v_D\,\sigma^3\tau^0(\sin k_y s^1 - \sin k_x s^2)\nonumber\\
&&{}+ \big[m + b(2-\cos k_x - \cos k_y)\big]\,\sigma^0\tau^2 s^3 \nonumber\\
&&{} + t_\textrm{a,1}\, \sigma^1 \tau^0 s^0 + t_\textrm{a,2} (\cos k_z\,\sigma^1 - \sin k_z\,
\sigma^2)\,\tau^0 s^0 \nonumber\\
&&{}+ t_\textrm{b,1}\, \sigma^1\tau^1 s^0
+\delta_1\, \sigma^2\tau^2 s^0 \nonumber\\
&&{}+ t_\textrm{b,2}\, (\cos k_z\, \sigma^1 - \sin k_z\, \sigma^2 )\,\tau^1 s^0
\nonumber\\
&&{}+ \alpha \sin k_z\, \sigma^3\tau^0(s^1+s^2),
\label{eq:tb_Hamiltonian}
\end{eqnarray}
In agreement with the low-energy model of the main part of the paper, this Hamiltonian preserves time-reversal symmetry 
and has inversion symmetry for $\delta_1=0$. In addition, the
Bloch Hamiltonian for $\alpha=0$ is invariant under a $\pi$ rotation around the $z$ axis, 
with the rotation operator given by $R_z(\pi)=i\sigma^0\tau^0
s^z$, and $(k_x,k_y,k_z)\rightarrow (-k_x,-k_y,k_z)$. To break this symmetry explicitly we have introduced
a rotational-symmetry breaking term parameterized by $\alpha$.

We find
that around $k_x=k_y=0$ and for $\alpha=0$ the
Hamiltonian in Eq.~\eqref{eq:tb_Hamiltonian} is identical to the effective model
discussed in the main part. As a consequence, both models have qualitatively the same phase
diagrams including Weyl-semimetal and STI phases (see Fig.~2 of the main part). In particular, for the gapped phases we can now calculate the
$\Z_2$ invariants directly. For this, we either restore inversion symmetry
and use the parities of the eigenstates at the six time-reversal invariant
points in the BZ~\cite{FuK07}, or we use a Wannier-center formulation of the
topological invariants~\cite{YQB11}. Both methods yield the same results. As
expected, our calculations confirm the invariants of the STI phases to be
$1;(000)$. 

\begin{figure}[t]\centering
\includegraphics[width=1.0\columnwidth]
{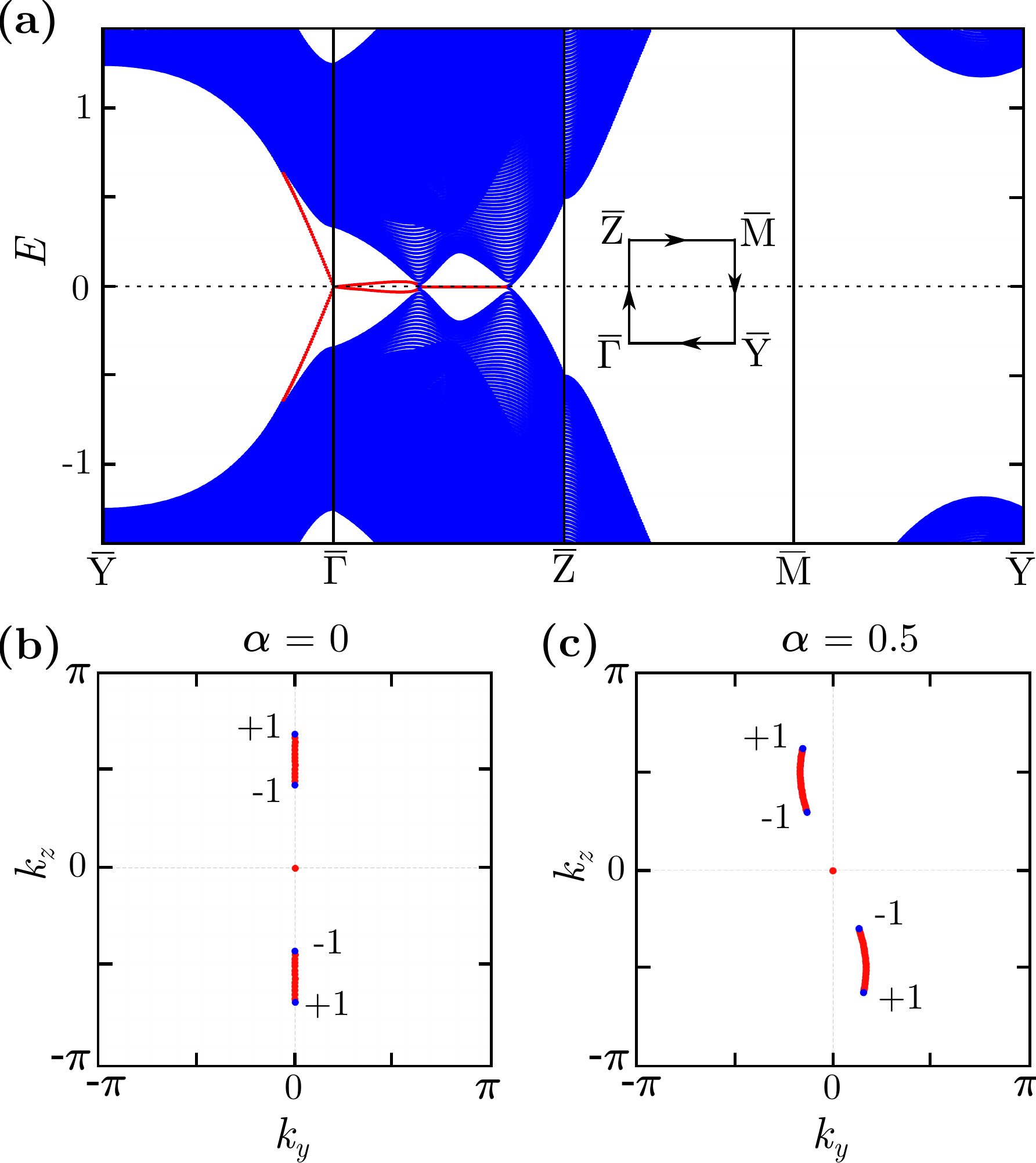}
\caption{(color online) Band structure and $E_F=0$ surface Fermi surfaces for the
WTI multilayer tight-binding model in a slab geometry with $v_D=m=b=t_\mathrm{a,1}=t_\mathrm{a,2}=1$,
$t_\textrm{b,1}=t_\textrm{b,2}=0.7$, and $\delta_1=0.5$ (Weyl
semimetal phase). States localized to the surfaces are highlighted in red. (a)
Band structure along high-symmetry lines of the surface BZ. Note the
surface Dirac cone around $\bar{\Gamma}$, and the surface Fermi arcs and bulk
Weyl cone projections along $\bar{\Gamma}\bar{Z}$. (b) Surface Fermi surface for
$\alpha=0$ (with rotational symmetry). (c) Surface Fermi surface for $\alpha=0.5$ (wihtout
rotational symmetry). The topological charges of the Weyl
nodes are also indicated.}
\label{fig:bandstructure_and_Fermi_surface}
\end{figure} 

Let us now investigate the surface features of our system in the Weyl semimetal phase. For
that, we carry out an inverse Fourier transformation of the
Hamiltonian in Eq.~\eqref{eq:tb_Hamiltonian} with respect to $k_x$ and study the
resulting mixed position-momentum space Hamiltonian with
open boundary conditions in the $x$ direction. This setup corresponds to a slab
geometry with two surfaces representing the dark surfaces of the underlying 
WTI layers. Energies and eigenstates are obtained by exact
numerical diagonalization.

In the presence of rotational symmetry,
the surface Fermi surface with respect
to a Fermi energy at $E_F=0$ contains four isolated bulk states along the $k_z$
axis which correspond to the surface projections of the four bulk Weyl nodes
[see Fig.~\ref{fig:bandstructure_and_Fermi_surface}(b)]. The isolated bulk
states are connected pairwise by doubly degenerate Fermi arcs of states
localized at the two surfaces, as expected for a Weyl semimetal. 
The vanishing curvature of the Fermi arcs is a consequence of rotational symmetry.  
Moreover, we find a pair of isolated, doubly degenerate surface states
pinned to the $\bar{\Gamma}$ point of the surface BZ. These states belong to the vertex
of a surface Dirac cone which happens to coincide with the Fermi level. Hence, our system is
yet another 
realization of a time-reversal invariant Weyl semimetal with coexisting Fermi arcs and
Dirac cones at its surfaces~\cite{LKB17,TSG17}.

The dispersion of the energy states along high-symmetry lines of the surface BZ
is illustrated in Fig.~\ref{fig:bandstructure_and_Fermi_surface}(a). The
projections of the bulk Weyl cones are clearly visible along the $k_z$
direction as well as the connecting Fermi arcs. In the $k_y$
direction we see the linear dispersion of the surface Dirac cone.
Its dispersion along the $k_z$ axis is extremely flat and
terminates at one of the Weyl nodes. 
When rotational symmetry is broken ($\alpha\neq 0$) the Weyl node projections
move away from the $k_z$ axis and the surface Fermi arcs are no longer straight
lines, as can be seen in Fig.~\ref{fig:bandstructure_and_Fermi_surface}(c).

\section*{B: Tight-binding model for topological crystalline insulator multilayers}

Following the same procedure as in the previous section, we 
now construct
a tight-binding model based on the TCI multilayer Hamiltonian of Eq.~(2) in the main part of the paper. To achieve this, we first perform the following substitutions: $k_x\rightarrow \sin k_x$, $k_y\pm d \rightarrow \sin (k_y\pm d)$. As before, this leads to a multitude of Weyl points. Another substitution, namely
\begin{eqnarray}
m\;\sigma^0 \tau^3 s^3 &\rightarrow&
[m + b(2 - \cos k_x)]\sigma^0 \tau^3 s^3\nonumber\\
&&{} - \frac{1}{2}b\cos(k_y+d)\,\sigma^0 (\tau^3 + \tau^0)s^3\nonumber\\
&&{} - \frac{1}{2}b\cos(k_y-d)\,\sigma^0 (\tau^3 - \tau^0)s^3
\end{eqnarray}
finally yields a tight-binding model with only four Weyl nodes, and whose low energy theory is thus identical to the Hamiltonian of Eq.~(2) in the main part. The Bloch Hamiltonian reads
\begin{eqnarray}
H(\mathbf{k}) &=&
\frac{v_D}{2}\,\sigma^3(\tau^0+\tau^3)[\sin (k_y+d) s^1 - \sin k_x s^2]\nonumber\\
&&{}+ \frac{v_D}{2}\,\sigma^3(\tau^0-\tau^3)[\sin (k_y-d) s^1 - \sin k_x s^2]\nonumber\\
&&{}+ \frac{1}{2}\big(m + b[2-\cos k_x - \cos (k_y+d)]\big)\,\sigma^0(\tau^3+\tau^0) s^3 \nonumber\\
&&{}+ \frac{1}{2}\big(m + b[2-\cos k_x - \cos (k_y-d)]\big)\,\sigma^0(\tau^3-\tau^0) s^3 \nonumber\\
&&{} + t_1\, \sigma^1 \tau^0 s^0 
+ t_2 (\cos k_z\,\sigma^1 - \sin k_z\,
\sigma^2)\,\tau^0 s^0 \nonumber\\
&&{}+ \alpha \sin k_z\, \sigma^3\tau^0(s^1+s^2).
\label{eq:TCI_tight_binding}
\end{eqnarray}
Similar to the tight-binding model for the WTI multilayer, this model is invariant under a $\pi$ rotation about the $z$ axis (for $\alpha=0$). For this reason, we have incorporated an additional rotational-symmetry breaking term parameterized by $\alpha$.

First of all, we compute the phase diagram of the tight-binding model by keeping track of all the gap-closing and reopening transitions. These happen approximately at the same points and for the same parameter values as for the low-energy model discussed in the main part. In particular, we find gapless Weyl semimetal phases and also insulating phases (see Fig.~4 of the main part). For the insulating phases we again calculate the four $\Z_2$ invariants explicitly. We find that they are in agreement with the values determined in the main part solely based on adiabaticity arguments, i.e., the invariants of the WTI phases are $0;(001)$ whereas those of the trivial phase are $0;(000)$. Furthermore, the Weyl-semimetal phase has four Weyl nodes of charge $\pm 1$.

\begin{figure}[t]\centering
\includegraphics[width=1.0\columnwidth]
{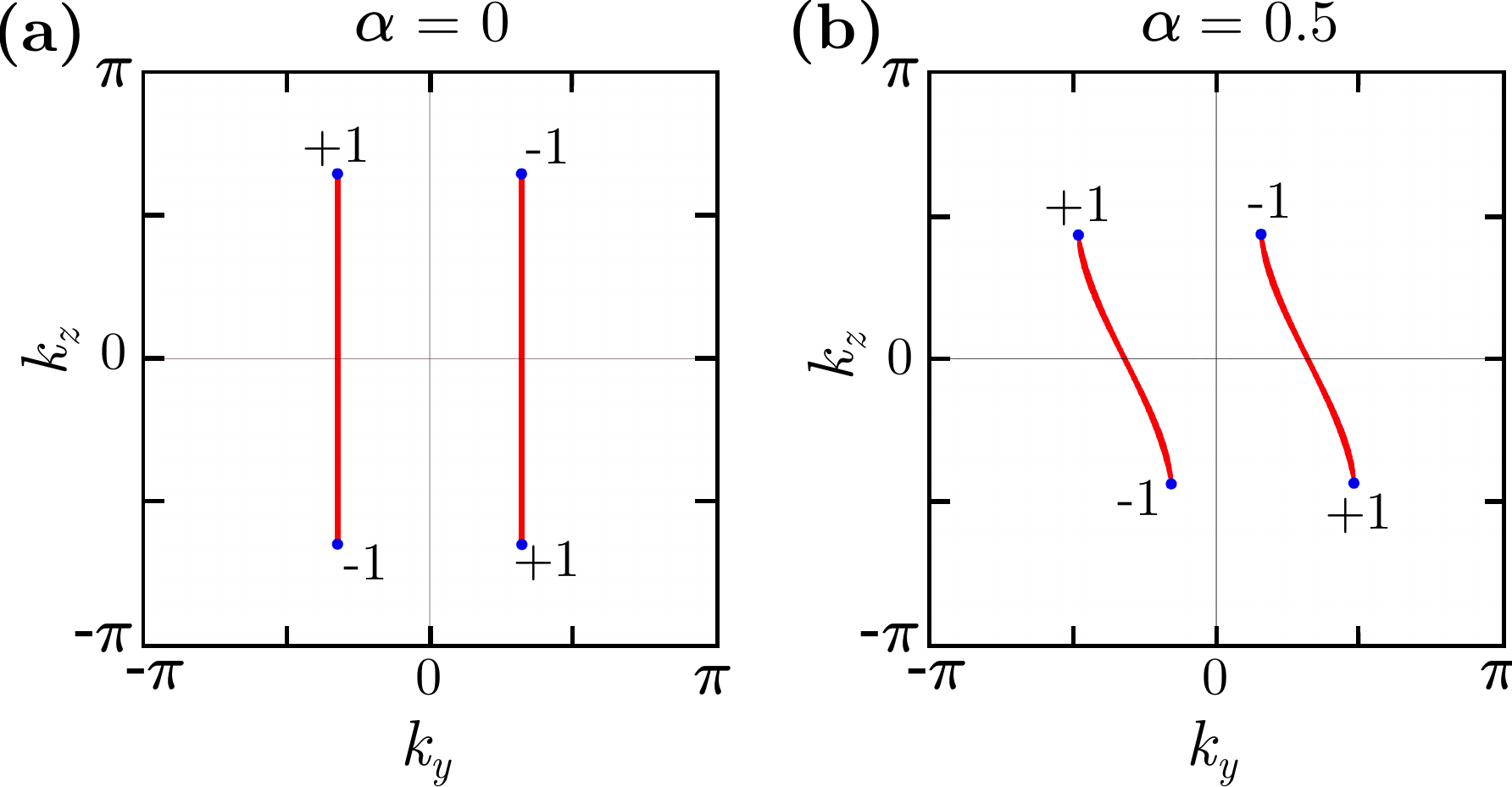}
\caption{(color online) $E_F=0$ surface Fermi surfaces for the
TCI multilayer tight-binding model with $v_D=b=1$,
$t_1=t_2=m=0.5$ and $d=1$ (Weyl
semimetal phase). The surface projections of the bulk Weyl nodes are highlighted in blue: (a) with rotational symmetry ($\alpha=0$). (b) broken rotational symmetry ($\alpha=0.5$). The topological charges of the Weyl nodes are also indicated.}
\label{fig:bandstructure_and_Fermi_surface_TCI}
\end{figure} 

To investigate the structure of the Fermi arcs in the Weyl-semimetal phase, we determine the (100) surface Green's function of the tight-binding model in Eq.~\eqref{eq:TCI_tight_binding} for a semi-infinite slab~\cite{SSR85}. This allows us to calculate the spectral function $A(\mathbf{k},E) = -1/2\pi\,\mathrm{Im}\lbrace \mathrm{Tr}[G_s(\mathbf{k},E)]\rbrace$ at the Fermi level $E=E_F$. Since our Hamiltonian is bilinear, the spectral function is sharply peaked whenever there is an eigenstate of the Hamiltonian at the Fermi level. Hence, we can use the spectral function to determine the surface Fermi surface of the semi-infinite slab. 

We show our results in Fig.~\ref{fig:bandstructure_and_Fermi_surface_TCI}. In the rotation-symmetric case [Fig.~\ref{fig:bandstructure_and_Fermi_surface_TCI}(a)], two straight lines of surface states connect the surface projections of the Weyl nodes. By breaking rotation symmetry, as shown in Fig.~\ref{fig:bandstructure_and_Fermi_surface_TCI}(b), the Weyl nodes are displaced and the Fermi arcs acquire a finite curvature.

\end{document}